\documentclass[10pt, journal,twoside]{IEEEtran}

\usepackage{graphicx}
\usepackage{amsmath}
\usepackage{balance}
\usepackage{amssymb}
\usepackage{cases}
\usepackage[font=footnotesize,labelfont=bf]{caption}
\usepackage{caption}
\usepackage{color}
\usepackage{multirow} 
\usepackage{booktabs}
\usepackage{algorithm}
\usepackage{algorithmic}
\usepackage[marginal]{footmisc}

\makeatletter
  \newcommand\figcaption{\def\@captype{figure}\caption}
  \newcommand\tabcaption{\def\@captype{table}\caption}
\makeatother
\usepackage{setspace}
\usepackage{bm}
\usepackage{setspace}
\usepackage{xcolor}
\usepackage{makecell}
\usepackage{booktabs}
\usepackage{multirow}
\usepackage{tabularx}
\usepackage{colortbl}
\usepackage{subcaption}
\usepackage{flushend} 

\begin{document}
\title{Multi-Party Data Pricing for Complex Data Trading Markets: A Rubinstein Bargaining Approach}

\author{Bing Mi, Zhengwang Han, Kongyang Chen
\IEEEcompsocitemizethanks{
\IEEEcompsocthanksitem{Bing Mi is with Guangdong University of Finance and Economics, Guangzhou 510320, China.}
\IEEEcompsocthanksitem{Zhengwang Han and Kongyang Chen are with Guangzhou University, Guangzhou 510006, China.}
}}

\IEEEtitleabstractindextext{
  \begin{abstract}
With the rapid development of Internet of Things (IoT) and artificial intelligence technologies, data has become an important strategic resource in the new era. However, the growing demand for data has exacerbated the issue of \textit{data silos}. Existing data pricing models primarily focus on single factors such as data quality or market demand, failing to adequately address issues such as data seller monopolies and the diverse needs of buyers, resulting in biased pricing that cannot meet the complexities of evolving transaction scenarios. To address these problems, this paper proposes a multi-party data pricing model based on the Rubinstein bargaining model. The model introduces buyer data utility indicators and data quality assessments, comprehensively considering factors such as the utility, accuracy, and timeliness of data sets, to more accurately evaluate their value to buyers. To overcome the limitations of single-factor models, this paper innovatively introduces the buyer data set satisfaction indicator, which reflects the overall satisfaction of buyers with data sets by integrating both data utility and quality assessments. Based on this, the model uses the Rubinstein bargaining model to simulate the pricing process between multiple sellers and multiple buyers, yielding pricing results that better align with market demands. Experimental results show that the proposed model effectively addresses the pricing imbalance caused by data monopolies and demonstrates good applicability and accuracy in multi-seller, multi-buyer transaction environments. This research provides an effective pricing mechanism for complex data trading markets and has significant theoretical and practical value in solving pricing issues in actual data transactions.
  \end{abstract}
}

\maketitle
\IEEEdisplaynontitleabstractindextext
\IEEEpeerreviewmaketitle

\section{Introduction}
In recent years, with the continuous development of the big data era, the demand for data has been steadily increasing, and data has become a critical strategic resource in the new era. With the rapid advancement of internet technologies, artificial intelligence, and Machine Learning as a Service (MLaaS), various derivative applications have emerged. However, despite the burgeoning demand for data, the issue of \textit{data silos} has become increasingly severe. Government departments and related enterprises store vast amounts of relevant data in their daily operations, data that can play a crucial role in the smart upgrade of industries. However, due to the lack of corresponding pricing and transaction mechanisms, these data cannot be reasonably circulated in the market. In the field of artificial intelligence, many practitioners are limited by the lack of high-quality and relevant datasets, which hinders the improvement of model accuracy. As a result, the trading of datasets for machine learning models has garnered increasing attention, and potential data buyers in the market urgently need reliable channels to obtain stable and trustworthy datasets. Therefore, how to scientifically and rationally price trading datasets has become an urgent issue that needs to be addressed.

Currently, traditional pricing model research is primarily based on cost, quality, or market demand. These traditional pricing models have been widely applied and validated in conventional goods and services. However, in the field of data pricing, the unique characteristics of data products—such as non-consumability, ease of replication, and easy dissemination—render existing experience and models not entirely applicable. Moreover, most of the current research models focus on single-factor influences. For instance, quality-based evaluation models mainly discuss the impact of a single data quality factor on data pricing. Most of the current transaction models based on game theory consider competitive monopolistic market types and discuss buyer-seller alliances to a lesser extent. Given that machine learning models require large quantities of multi-category data for training, the transactions in such scenarios should not be limited to single-to-many or many-to-one situations. In current research on data pricing transactions, there is limited exploration of the potential for monopolistic oligopoly situations, and how to reasonably price monopolized datasets remains one of the key challenges.

This paper focuses on the data pricing scenario between multiple data sellers and buyers, analyzing the issue of data seller monopolies in current pricing models. From the perspective of the data buyer's actual needs, the paper introduces the buyer's data utility index. This index allows buyers to intuitively understand whether the data provided by the seller meets their model requirements. In practical applications, in addition to the utility function of the dataset itself, factors such as accuracy and timeliness should also be considered. Therefore, the proposed data pricing model not only considers the influence of buyer data utility on data pricing but also integrates four data quality indicators—accuracy, completeness, consistency, and timeliness—into the pricing process. This paper innovatively introduces the buyer dataset satisfaction index, which combines buyer data utility and data quality indicators, and uses the Rubinstein bargaining model to simulate actual pricing problems. This makes the final pricing result more aligned with the buyer's actual needs and the current market evaluation of the seller's dataset.

The data pricing model proposed in this paper innovatively introduces the buyer dataset satisfaction index, which combines the buyer's data utility and data quality evaluations of each seller's dataset. This avoids the risk of transaction failure caused by data monopolies. The pricing model then facilitates bargaining between buyer alliances and each seller to ultimately determine the pricing of the dataset. Previous methods generally only support single seller-multiple buyer or multiple seller-single buyer transaction forms and do not fully consider the impact of data monopolies. These models fail to meet the increasingly complex pricing scenarios. In contrast to prior approaches, the model proposed in this paper is the first to address the risk of data monopoly in multi-seller transaction scenarios and provides specific solutions. Additionally, this model better suits multi-seller multi-buyer pricing scenarios, expanding the context of data pricing while fully considering the real needs of both buyers and sellers, offering greater generality and applicability.

The primary contributions of this study are as follows:
\begin{enumerate}
\item \textit{Addressing Data Seller Monopolies:} This paper is the first to investigate the impact of data seller monopolies on pricing in multi-seller multi-buyer scenarios. It highlights how monopolistic control over data sets can skew pricing and transaction outcomes, and proposes a specific pricing model that accounts for the unique gains from such monopolies. This approach helps ensure that pricing reflects the true value of monopolistic data in the market.
\item \textit{Introduction of the Buyer Dataset Satisfaction Index:} A significant innovation in this paper is the introduction of the Buyer Dataset Satisfaction Index, which integrates both the buyer's data utility and the dataset quality. This index is calculated using a weighted combination of factors such as accuracy, completeness, consistency, and timeliness. It provides a more accurate reflection of a buyer's satisfaction with a dataset, ensuring that the pricing process better aligns with the buyer's actual needs and expectations.
\item \textit{Incorporation of Utility and Quality in the Bargaining Model:} Unlike previous pricing models that focus primarily on individual factors like data utility or quality, this paper introduces both data utility and quality indicators into the bargaining process. By using the Rubinstein bargaining model with the Buyer Dataset Satisfaction Index, the paper presents a more comprehensive and realistic approach to data pricing. This method ensures that both the utility of the dataset for the buyer and its quality characteristics are properly accounted for in the final price, leading to more credible and fair pricing outcomes.
\end{enumerate}

The remainder of the paper is structured as follows: Section~\ref{sec:related} elaborates on related work. Section~\ref{sec:formation} describes the problem formation. Section~\ref{sec:method} presents our research methodology. Section~\ref{sec:experiment} provides a detailed explanation of the experimental results. Finally, Section~\ref{sec:conclusion} concludes the paper.

\section{Related Work}
\label{sec:related}
In the research on data pricing~\cite{ref1}, the unique characteristics of data products—namely, their extremely low replication and distribution costs—significantly reduce the effectiveness of traditional pricing strategies for these products. The core idea to address this issue is to link the price of a product to its value, correlating the product's pricing with the buyer's perceived importance of the information. For instance, different versions can be created~\cite{ref11} to meet the varying needs of buyers. Paid predictive APIs represent a rapidly growing industry and form an essential component of machine learning as a service~\cite{ref50}. Based on this, pricing strategies can be determined by two key factors: first, the importance of the information being sold, and second, the degree to which different buyers value the product. The method of version control has already been extensively researched in the context of data products, such as through relational data set queries~\cite{ref12}. In the process of data pricing and trading, besides the challenges posed by the inherent characteristics of data, it is essential to ensure smooth transactions and maximize the interests of all parties involved. In this regard, Kushal et al.~\cite{ref10} defined the ideal attributes that data products should possess during transactions. Ren et al.~\cite{ref59} summarized the data pricing models in various types of data markets. Zhang et al.~\cite{ref64} conducted a survey on data pricing models within game theory and auctions. Pei et al.~\cite{ref61} were among the first to explore the governance of data assets in collaborative artificial intelligence. Furthermore, Pei~\cite{ref20} proposed a set of standard guidelines that all parties involved in data transactions should follow. These guidelines cover principles such as integrity, fairness, no arbitrage, profit maximization, privacy protection, computational efficiency, and rationality of participants.

With the growing prominence of large models and deep learning, which rely heavily on vast amounts of high-quality datasets, Yang et al.~\cite{ref62} analyzed the impact of data quality on big data analytics from a data science perspective and defined a utility function for data quality. In the context of machine learning marketplaces, recent research and practices have successfully commoditized data in various ways. Data markets sell data either directly or indirectly, yet current pricing mechanisms vary due to different application scenarios~\cite{ref68, ref72, ref70}. Data trading markets can currently be classified according to their pricing mechanisms and the types of data sold.

\textit{Pricing Based on Raw Data:} 
Pricing strategies based on datasets allow data markets to sell datasets and grant buyers access to raw datasets, such as those from Dawex, Twitter, Bloomberg, Iota, and SafeGraph. Traditional approaches price datasets as indivisible units, with inherent attributes, such as quantity, serving as key factors determining the price.

\textit{Query-Based Pricing:} 
Query-based pricing models enable data buyers to purchase datasets of interest. Koutris et al.~\cite{ref27} were the first to propose query-based data pricing, establishing a corresponding transaction framework that sets prices for arbitrary queries while ensuring the no-arbitrage principle. In subsequent work, Koutris et al.~\cite{ref18} redesigned and improved the query market framework, addressing the limitations of simple query pricing. They optimized the complex computations involved in processing large numbers of SQL queries into an integer linear programming problem, enhancing computational efficiency. In addition to view-based pricing, Tang et al.~\cite{ref28} introduced a minimum-source pricing model for tuples. Shen et al.~\cite{ref29} conducted research on personal data attributes and established a personal data pricing platform from the tuple perspective. Niu et al.~\cite{ref66} proposed a context-aware dynamic pricing mechanism with a low-price constraint, which maximizes cumulative revenue by setting reasonable prices for sequential queries and achieves efficient online optimization.

\textit{Model-Based Pricing:} 
Many scenarios today require machine learning models, and many users or companies do not build machine learning models from scratch but rather purchase pre-trained models. Chen et al.~\cite{ref30} proposed a machine learning model marketplace based on the no-arbitrage principle and the revenue-maximization principle, where model owners sell multiple versions of machine learning models to different buyers. Liu et al.~\cite{ref31} introduced an end-to-end model marketplace that accounts for privacy compensation for data sellers and the needs of model buyers. Lin et al.~\cite{ref37} used knowledge graph methods to specifically assess the value of data in models and provided a calculation approach for this evaluation.

\textit{Pricing Based on Data Quality:} 
In data transactions, higher-quality data holds greater value for buyers. Determining the quality of data products is a key focus in data markets. Regarding data quality evaluation, Heckman et al.~\cite{ref32} identified a series of factors for assessing dataset quality and proposed a linear model for pricing based on data quality. Yu et al.~\cite{ref33} explored pricing methods based on multiple quality dimensions of data in a monopoly market and designed a transaction model involving a data market and several data buyers. Ding et al.~~\cite{ref63} developed a fair data pricing evaluation mechanism aimed at meeting the demands of both supply and demand sides. In addressing data quality issues in data markets, they integrated key factors such as accuracy, completeness, consistency, and timeliness into the pricing of data.

\textit{Pricing Based on Privacy Protection:} 
During data transactions, data buyers may infer the privacy of data providers based on the characteristics of the data itself or through model-based inferences. Jiang et al.~\cite{ref34} proposed a market framework for trading private data generators under differential privacy using GANs. In this framework, the trading platform charges query fees to data buyers while compensating data providers for privacy. Li et al.~\cite{ref35} developed an incentive-compatible mechanism from the perspective of data providers to price their data, investigating the privacy issues that may arise when data owners release data with unclear usage purposes, which could be distributed by third-party users. Yu et al.~\cite{ref41} used a matching-based Markov decision process to model multi-round data transactions involving gradually disclosed information, introducing a social welfare-optimized data pricing mechanism to identify the best pricing strategy. Feng et al.~\cite{ref75} proposed a privacy-aware personalized data transaction method based on contract theory, offering a set of optimal contracts with varying levels of privacy protection and data transaction prices for self-interested data owners.

\textit{Pricing Based on Task:} 
When pricing data labels, "golden tasks" can be introduced to incentivize pricing. Shah et al.~\cite{ref36} developed a golden task pricing method, which mixes golden tasks—tasks where the data buyer knows the answers and their purpose—with regular tasks, and then assigns them to workers. Since workers cannot distinguish between golden tasks and regular tasks, this method can be used to assess workers' performance and effectively incentivize them to provide accurate labels.

However, current research on data pricing primarily focuses on raw data or query-based approaches. These pricing methods often consider only specific characteristics of the data itself or the demand-supply dynamics of the buyer-seller relationship, leading to an inability to accurately measure the true value of data. Additionally, many existing data pricing models are based on competitive scenarios, where buyers select a seller from multiple options. However, these models do not address scenarios where data sellers possess monopolistic datasets. Data quality, a key factor influencing the value of data, plays an essential role in data-driven model training, such as in machine learning and large model applications. High-quality data is crucial for improving model accuracy. Currently, data quality evaluation tends to focus on the data itself or the buyer's needs, with some approaches using Shapley values to assess the utility of data in model pricing. However, these methods are often limited and fail to integrate the buyer's utility with data quality characteristics in their evaluations.

\section{Problem Formulation}
\label{sec:formation}
As a key element in unlocking the value of data products, data pricing has gained growing attention. Current research on data pricing largely focuses on single factors such as cost, quality, or data utility, which fails to meet the increasingly dynamic market demands.

In multi-party data pricing scenarios, due to the non-competitiveness within both the seller's and buyer's internal groups, existing pricing methods are inadequate for evaluating the price of seller data sets in joint transactions. In multi-party buy-sell scenarios, data sellers $S=\{S_1,S_2, \cdots,S_n\}$ form a seller alliance and provide data sets $D=\{D_1,D_2, \cdots,D_n\}$ along with reserve prices $v=\{v_1,v_2,\cdots,v_n\}$, where $t_{C_i}$ represents the dataset of category $C_i$ within data set $D$. Data buyers $B=\{B_1,B_2,...,B_m\}$ form a buyer alliance and provide their required models $M=\{M_1,M_2,...,M_m\}$ along with their maximum budgets $u=\{u_1,u_2,...,u_m\}$. The transaction platform matches the buyers' and sellers' needs, after which both parties price the data sets for the transaction. The key issues that need to be addressed in this transaction scenario are as follows:

\begin{figure*}[!ht]
    \centering
    \includegraphics[width=1\linewidth]{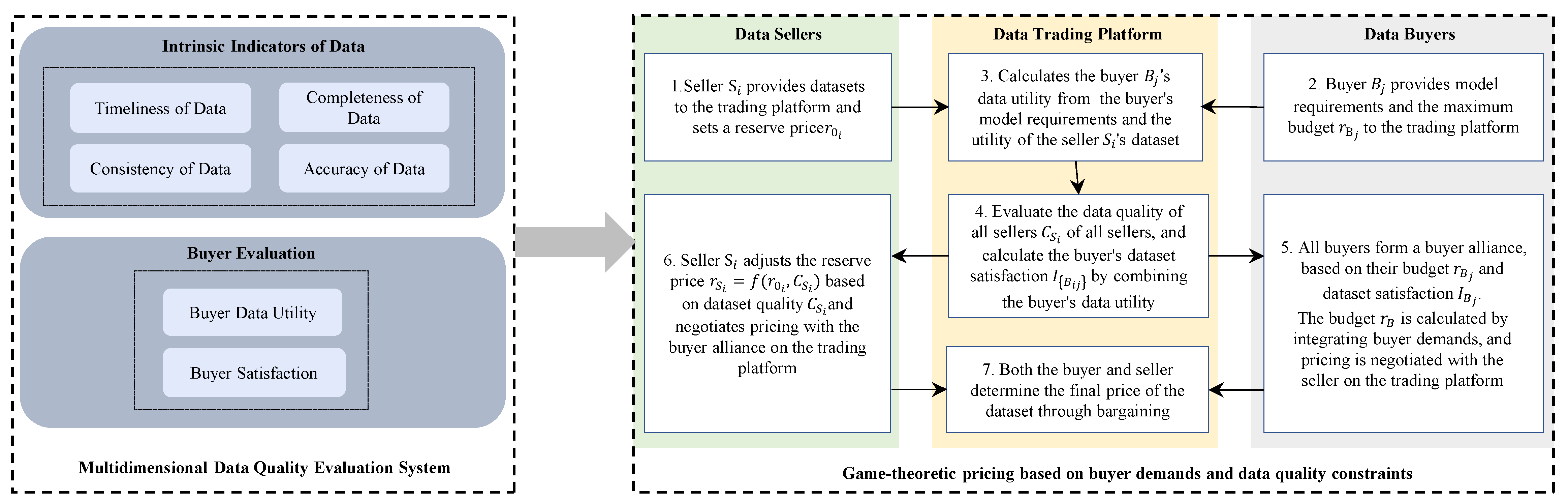}
    \caption{System framework.}
    \label{fig:overview}
\end{figure*}

\begin{enumerate}
\item Assuming that data seller $S_1$  holds a larger share of critical or monopolistic data within the alliance, i.e., when $t_{C_j} \subset  D_1$ and $t_{C_j} \cap D_i= \emptyset$ for all $i=\{1,2,...,n\}$, how can the monopolistic bidding advantage be assessed? \item When the buyer alliance $B=\{B_1,B_2,...,B_m\}$ prices data set $D_i$, how can the budget for each individual buyer $B_j$ for $D_i$  be determined? 
\item How can the pricing model comprehensively incorporate factors such as the attributes of the data set $D=\{D_1,D_2,...,D_n\}$ (e.g., cost, quality) along with the buyer's utility to reasonably price the data set? 
\end{enumerate}

\section{Our Method}
\label{sec:method}
In this section, we address the challenges in multi-party joint data pricing scenarios by proposing a buyer data set satisfaction metric based on buyer data utility and data quality, and using the Rubinstein model to price data sets reasonably. 

\subsection{Overview}
Despite the longstanding research on data pricing, a universal pricing model has yet to emerge from both an economic and big data application technology perspective. This paper addresses the issues of insufficient consideration of factors in current pricing mechanisms, the deviation of dataset value from its true value, and unclear data utility. We propose a multi-factor comprehensive Rubinstein pricing model. Specifically, the model first considers that each buyer has different model requirements and identification needs, introducing a buyer's data utility assessment to calculate the overall utility value of each dataset for individual buyers. The model then evaluates dataset quality and combines both buyer data utility and data quality to determine buyer satisfaction, reflecting the buyer's interest in the dataset. During the bargaining process, multiple buyers form a buyer alliance, and based on their budgets and data utility, the alliance determines the reserve price for each seller and the buyer alliance's dataset satisfaction. Sellers dynamically adjust dataset prices according to dataset quality. Subsequently, both parties negotiate and finalize the dataset price. Experiments validate the effectiveness of this pricing model in handling multi-party joint data buying and selling pricing problems. 
The system framework of the Rubinstein pricing model based on multiple factors is shown in Figure~\ref{fig:overview}.

\subsection{Buyer Data Utility}
In real-world data trading market environments, data sets often experience category monopolies, which pose significant challenges to buyers who require specific categories of data sets to meet their training needs. Suppose there exists a seller set $S=\{S_1,S_2,...,S_n\}$ holding data sets $D=\{D_1,D_2,...,D_n\}$, and a buyer set $B=\{B_1,B_2,...,B_n\}$, where each buyer forms a coalition to jointly purchase data points that best meet their needs. If a specific category is monopolized by a particular seller, and this seller sets a high price for that category, while the buyers requiring this category are relatively few and lack sufficient budget, the coalition purchase may fail. For instance, if buyer $B_i$ requires data from a specific category in the data set $D$ for their model training task, and this category is monopolized by seller $P_i$, despite Shapley values helping buyer $B_i$  avoid purchasing irrelevant data points, the high price set by seller $P_i$ may still negatively affect $B_i$'s ability to participate in the joint purchase, especially if other buyers do not need that particular category of data. This creates an unfair situation for buyers with specific needs and may prevent other buyers from obtaining the data they require. To address this issue, we introduce the concept of \textit{buyer data utility} in this paper.

\textbf{Buyer Data Utility $\xi_{i,j}$}:  This metric is defined as the overall utility that the data subset $D_i$  owned by seller $S_i$ provides to the model requirements of data buyer $B_i$. This utility needs to be assessed based on the specific model and identification requirements of the data buyer. The buyer's identification needs are used as a criterion for selecting a test set for the model or are provided by the buyer as the test set. Subsequently, the Shapley algorithm is applied to evaluate the buyer data utility $\xi_i$ for the entire set of sellers from the perspective of buyer $B_i$.

Through Algorithm~(\ref{alg:buyer}), we derive the buyer utility evaluation values $\xi$ for all buyers, which can be considered as the ability of the dataset to meet the specific needs of the buyer. This serves as a solid foundation for determining a reasonable price in the subsequent pricing problem. In the evaluation of buyer satisfaction metrics, the buyer utility evaluation value plays a critical role as the most important reference indicator. The method of buyer utility evaluation, from a practical perspective, ensures fairness for data buyers during the data transaction process and effectively prevents monopolistic data sellers, $P$, from setting unreasonable prices that could disrupt the transaction.

\begin{algorithm}[h]
\caption{Buyer Data Utility Evaluation Algorithm.}
\label{alg:buyer}
  \begin{algorithmic}[1]
   \STATE \textbf{Input:} Data set $D=\{D_1,D_2,...,D_n\}$, the model  $A$ required by the buyer, and the test set $D_t$ that meets the buyer's requirements.
   \STATE \textbf{Output:} Buyer data utility value $\xi$.
   \STATE \textbf{Initialize:} $\xi_i=0$ for $i=1,...,n$ 
   \FOR{$i \in \{ 1, \cdots, n \}$}
	   \FOR{Data set $U = \forall \{D-  {D_i}\}$}
		   \STATE $U_q = U \cup D_i$.
		   \STATE Marginal contribution $M=V(U_q, D_t, A) - V(U, D_t, A)$.
		   \STATE  $\xi_i= \frac{ |U|! \cdot  (|D|-|U|-1)!} {|D|!} \cdot M$.
	   \ENDFOR
	   \STATE Normalize $\xi_i$.
   \ENDFOR
  \end{algorithmic}
\end{algorithm}

\subsection{Data Quality Evaluation Model}
In the data transaction process, the quality of a dataset is a crucial pricing criterion. During the transaction, data quality directly affects the economic return of the data buyer. For instance, the timeliness of the dataset is a key indicator. If a dataset's timeliness is poor, models trained with it will likely underperform and fail to generate actual economic benefits. Therefore, purchasing a dataset with poor timeliness poses a high risk for the buyer. Current research on data quality evaluation is well-developed, and a reasonable and scientific data quality evaluation model should meet the following quintuple form:
$DQM = {D, T, R, C, W}$,
where $DQM$ refers to the data quality evaluation model.
$D$ represents the dataset being evaluated for quality.
$T$ denotes the quality metrics selected in the data quality evaluation model.
$R$ refers to the evaluation rules that need to be referenced when assessing data quality, with corresponding scores assigned when the rules are met. Different datasets may require different evaluation rules. In this study, five distinct gradation values are set for the scoring system.
$C$ represents the score assigned to the dataset after applying the evaluation rules $R$.
$W$ is the weighted factor in the data quality evaluation model for each of the evaluation metrics.

In the model, the evaluation metric for data quality, $T$, includes four indicators: \textit{timeliness, consistency, completeness, and accuracy}. There are two main reasons for selecting these four metrics. First, research on these metrics is well-established and comprehensive, making them some of the most commonly used factors in data quality pricing studies. Second, these metrics are intuitive and aligned with economic principles, directly correlating with the value of the data. Violating any of these indicators would result in significant economic losses. Other data quality indicators are either not directly related to data pricing or overlap with one of the four selected metrics. For example, accessibility does not directly affect data quality, and appropriate data volume overlaps with the completeness metric.

Regarding the evaluation rules $R$, this study has referenced and formulated basic rule descriptions along with five distinct gradation levels, as shown in Table~\ref{tab:data-quality}.

\begin{table*}[h]
  \centering
  \small
  \caption{Data Quality Evaluation Rules and Scoring Criteria~\cite{ref60}}
  \label{tab:data-quality}
\begin{tabular}{ccp{9cm}c}
\hline
\textbf{Data Quality Metric} & \textbf{Evaluation} & \textbf{Rule} & \textbf{Score} \\ \hline
\multirow{10}{*}{Timeliness}  & Excellent & The dataset contains the most current data, valid at the current time point, and can generate unexpected economic benefits & 1.2  \\
& Very Good & At least 80\% of the data is valid at the current time point and brings expected benefits to the buyer & 1 \\
& Satisfactory & At least 60\% of the data is valid at the current time point and brings benefits to the buyer & 0.6 \\
& Fair & Less than 40\% of the data is valid at the current time point, but still generates benefits for the buyer & 0.4 \\
& Unsatisfactory & The data is outdated and cannot be used, offering no benefits to the buyer & 0.2 \\
\hline
\multirow{10}{*}{Consistency} & Excellent & The data points in the dataset are strongly correlated with the features, are logically clear, and provide excellent problem-solving capabilities for the model & 1.2 \\
& Very Good & The data points are correlated with the features, the logic is clear, and the dataset provides good problem-solving capabilities for the model & 1 \\
& Satisfactory & The data points have some correlation, providing problem-solving capability for the model & 0.6 \\
& Fair & The data points lack clear correlation or contain logical errors but still provide some problem-solving capability for the model & 0.4 \\
& Unsatisfactory & The dataset lacks correlation, is logically chaotic, and cannot provide problem-solving capabilities for the model & 0.2 \\
\hline
\multirow{10}{*}{Completeness} & Excellent & The dataset is comprehensive and diverse, covering all the needs of data buyers & 1.2 \\
& Very Good & The dataset is complete and diverse, covering nearly all the needs of data buyers & 1 \\
& Satisfactory & The dataset is relatively complete and diverse, covering most of the needs of data buyers & 0.6 \\
& Fair & The dataset is incomplete and not very diverse, but still covers some of the needs of data buyers & 0.4 \\
& Unsatisfactory & The dataset is insufficient and lacks diversity, unable to meet the needs of data buyers & 0.2 \\
\hline
\multirow{10}{*}{Accuracy} & Excellent & The dataset is accurate and authentic, free of false data, and fully meets the buyer's needs & 1.2 \\
& Very Good & The dataset is mostly accurate, with at least 80\% of the data being valid, and can be used after minimal filtering to meet the buyer's needs & 1 \\
& Satisfactory & At least 60\% of the dataset is accurate and valid, and can be used after filtering, basically satisfying the buyer's needs & 0.6 \\
& Fair & Less than 40\% of the dataset is accurate, but it can be used after careful filtering and still meets the buyer's needs & 0.4 \\
& Unsatisfactory & The dataset is entirely inaccurate and cannot be used & 0.2 \\
\hline                                
\end{tabular}
\end{table*}

In the evaluation process of $R$, five levels of evaluation and scores are assigned for each metric. The highest score, 1.2, is assigned to the "Excellent" category, meaning that the highest-quality datasets in the market are rewarded with higher scores, allowing data sellers to gain more profit in subsequent transactions. In the data pricing model proposed in this paper, the evaluation is jointly conducted by all participating data buyers and the trading platform. The steps of this evaluation process are as follows:

\begin{enumerate}
\item The data buyer submits the required types of datasets and their corresponding timeline requirements to the trading platform. \item The trading platform investigates and verifies whether the dataset is authentic according to the evaluation rules, assigning an accuracy score $C_1$. 
\item The trading platform evaluates whether the dataset fully covers the buyer's required types and assigns a completeness score $C_2$. 
\item The trading platform assesses the consistency of the dataset through investigation, data analysis, and other methods, resulting in a consistency score $C_3$. 
\item The trading platform evaluates the timeliness of the dataset based on the timeline requirements submitted by the data buyer, assigning a timeliness score $C_4$. 
\end{enumerate}

Through these steps, the final four evaluation scores for the dataset are obtained. The overall score of the dataset is calculated as $C=1/4  \sum_{i=1}^4 C_i$ . This composite score indicates that all four metrics are weighted equally, implying that they are considered of equal importance. However, in real-world trading scenarios, the weights of data quality indicators should differ, reflecting varying levels of importance. This necessitates the introduction of weight parameters $W$ in the data quality evaluation model. After incorporating the weights, the comprehensive weighted score of the dataset is:
\begin{equation}
C = \sum_{i=1}^4 W_i C_i.
\end{equation}
noindent Where $0 \leq W_i \leq 1$ and $\sum_{i=1}^4 W_i =1$.

When the weight $W_i$ is zero, it indicates that the data quality factor does not need to be considered in subsequent data pricing. If the weights for all four metrics are equal, it implies that the importance of each metric is considered to be the same. In practical data pricing and trading, the weights should be differentiated based on the specific emphasis required by the buyer's needs. The derivation and calculation process of these weights will be provided in the next subsection.

\subsection{Buyer Data Set Satisfaction}
After the evaluation of the buyer’s data utility and the data quality model, the data utility $\xi_i$ for each buyer and the overall data quality score $C$ for all buyers can be obtained. Based on these two indicators, this paper introduces the concept of buyer data set satisfaction. In data pricing and trading, the most crucial reference for data buyers is whether the data set can meet their model recognition needs. That is, the buyer's data utility $\xi_i$ should be the primary consideration when assessing buyer satisfaction during the purchase. Subsequently, by integrating the four evaluation dimensions from the data quality model, the weighted buyer data set satisfaction metric can be derived.

\textbf{Buyer Data Set Satisfaction $I$}: This value is derived by applying weights $W$ to the buyer's data utility $\xi_i$ and the comprehensive data quality score $C$. The buyer data set satisfaction for buyer $B_i$  is given as follows:
\begin{equation}
     I_{B_i} = 
    \begin{cases}
        W_1 \xi_i+W_2 C_1+W_3 C_2+W_4 C_3+W_5 C_4  &  \mathrm{if~}   \xi_i>0    ,\\
        0                                                                                                     &  \mathrm{if~}   \xi_i=0.
    \end{cases}
\end{equation}

\noindent The weight $W$ plays a crucial role in aggregating multiple factors, as it significantly enhances the accuracy of the target influence during subsequent data pricing. The methodology for calculating weights has matured over time. In cases where there is limited reference data and the evaluation criteria are predominantly qualitative, the Analytic Hierarchy Process (AHP) is commonly used. In this paper, since the data is sparse and the indicators are based on qualitative analysis, the weight $W$is determined using the AHP method.

\begin{figure}[h]
    \centering
    \includegraphics[width=0.6\linewidth]{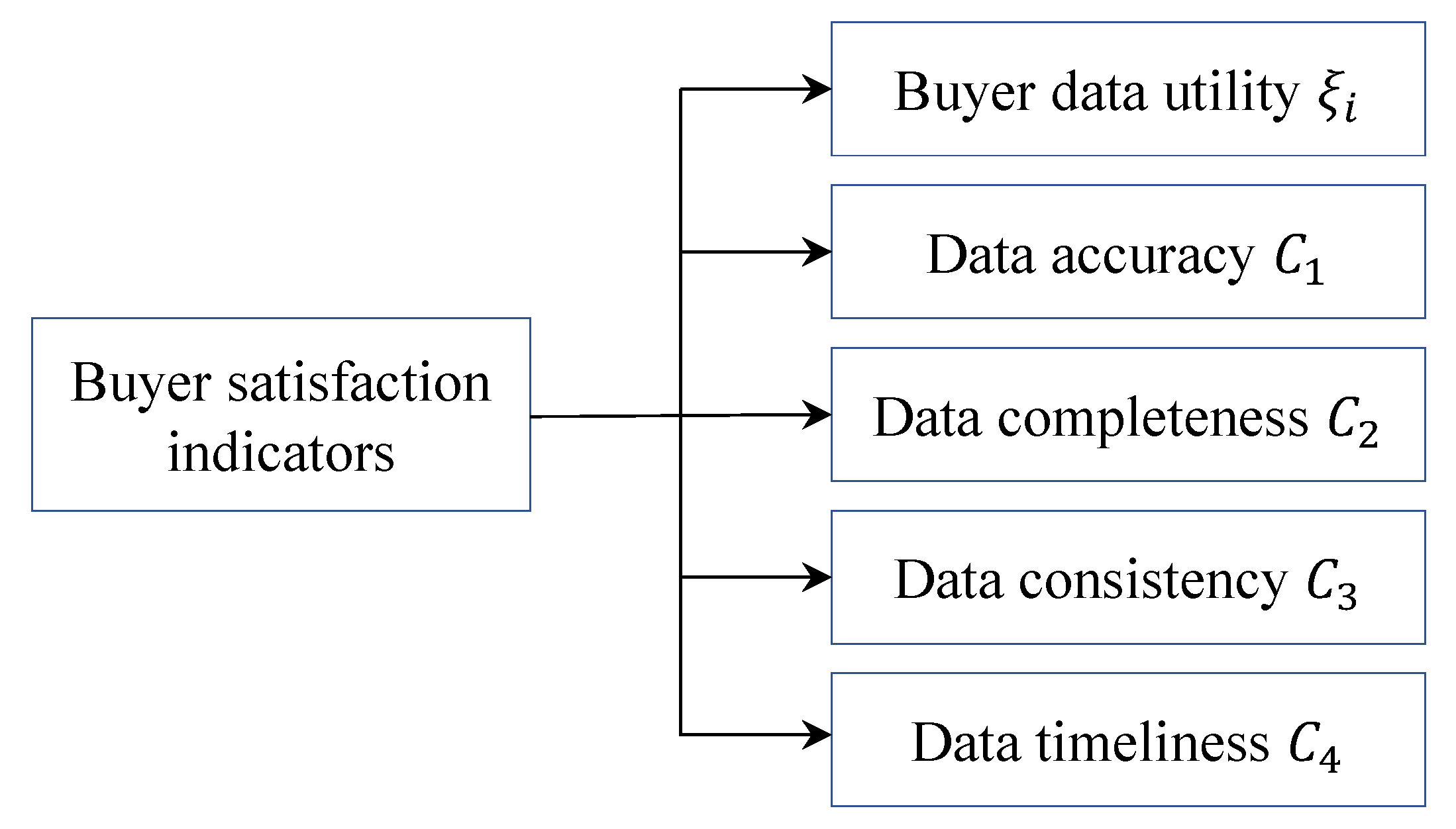}
    \caption{Hierarchy Structure of Buyer Data Set Satisfaction Indicators}
    \label{fig:buyer-satisfaction}
\end{figure}

\textit{Hierarchy Scale Values and Analytic Hierarchy Process (AHP): }
According to the steps of AHP, the first task is to analyze the hierarchical structure of the indicators corresponding to the weights. To determine the specific values of the weights $W$, it is essential to establish the hierarchy structure. The hierarchy includes buyer data utility $\xi_i$ , accuracy $C_1$, completeness $C_2$, consistency $C_3$, and timeliness $C_4$, which form a parallel structure as shown in Figure~\ref{fig:buyer-satisfaction}. For the sake of unified calculation and representation, $\xi_i$  will be treated as $C_5$ in subsequent calculations.

Using the three-point scale method $(0,1,2)$, a comparison matrix for the five indicators is established. This comparison matrix is assumed to be derived from a market survey conducted on actual data buyers. The comparison matrix is as follows:

\begin{equation}
C=(C_{ij})=
\begin{bmatrix}
C & C_1 & C_2 & C_3 & C_4 & C_5 & R_i  \\
C_1 & 1 & 2 & 2 & 2 & 0 & 7 \\
C_2 & 0 & 1 & 0 & 2 & 0 & 3 \\
C_3 & 0 & 2 & 1 & 2 & 0 & 5 \\
C_4 & 0 & 0 & 0 & 1 & 0 & 1 \\
C_5 & 2 & 2 & 2 & 2 & 1 & 9
\end{bmatrix}
\end{equation}

\noindent where the meaning of the scale values for the three-point scale method is shown in Table~\ref{tab:three-point}.
\begin{table}[h]
  \centering
  \small
  \caption{Meaning of the Three-Point Scale Values~\cite{ref60}}
  \label{tab:three-point}
\begin{tabular}{cc}
\hline
\textbf{Comparison of Element Importance}  & \textbf{Evaluation Value} \\ \hline
$C_i$ is more important than $C_j$ & 2 \\
$C_i$ and $C_j$ are equally important & 1 \\
$C_i$ is less important than $C_j$ & 0 \\
\hline                                          
\end{tabular}
\end{table}

Based on the values in the comparison matrix, the range method is used to construct a new comparison matrix. With $R_{max}=9$ and $R_{min}=1$, the range $R = R_{max} - R_{min} = 8$. During the process of constructing the comparison matrix, the value of $C_b$ is set to 9. Thus, the new comparison matrix $ H = C_b ^{ (R_i-R_j) / R }$ is:

\begin{equation}
\small
H= 
\begin{bmatrix}
C & C_1 & C_2 & C_3 & C_4 & C_5 & W \\
C_1 & 1 & 3 & 1.73 & 5.20 & 0.58 & 0.2693 \\
C_2 & 0.33 & 1 & 0.58 & 1.73 & 0.25 & 0.0950 \\ 
C_3 & 0.58 & 1.73 & 1 & 3 & 0.44 & 0.1647 \\
C_4 & 0.19 & 0.58 & 0.33 & 1 & 0.11 & 0.0516 \\
C_5 & 1.73 & 3.95 & 2.28 & 9 & 1 & 0.4195
\end{bmatrix}
\end{equation}

The Consistency Index (CI) is calculated as:
\begin{equation}
CI=\frac{\lambda_{max}-n} {n-1}.
\end{equation}
\noindent where $\lambda_{max}$  is the largest eigenvalue of the matrix, and $n$ is the order (dimension) of the matrix. In this case, the largest eigenvalue of matrix $\lambda_{max}=5.00437$, thus the consistency index is: $CI=\frac{5.00437 - 5} {5-1} = 0.0010927$.

Next, consistency verification is needed to check whether the computed matrix weights follow the rules. The verification method involves calculating the Random Consistency Ratio (CR):
\begin{equation}
CR=\frac{CI}{RI}.
\end{equation}
where $RI$  is the average random consistency index. According to Table~\ref{table:RI}, the average consistency index  $RI$  is 1.12.

\begin{table}[h]
  \centering
  \small
  \caption{Average Random Consistency Index Reference Values}
  \label{table:RI}
\begin{tabular}{cccccccccc}
\hline
n	&1	&2	&3	&4	&5	&6	&7	&8	&9 \\
RI	&0	&0	&0.52	&0.89	&1.12	&1.26	&1.36	&1.41	&1.45 \\ \hline
\end{tabular}
\end{table}

Substituting $CI=0.0010927$ and $RI=1.12$ into the formula: $CR=\frac{CI}{RI}=\frac{0.0010927}{1.12}=0.00098214$. Since CR is less than 0.1, the comparison matrix passes the random consistency check. Thus, the matrix weights are as follows:
\begin{equation}
W=(W_1,W_2,W_3,W_4,W_5)=(0.27, 0.10, 0.16, 0.05, 0.42).
\end{equation}
By substituting these weight values into the formula for buyer satisfaction $I_{B_i}$, the buyer satisfaction function is: 
$ I_{B_i} = 0.42\xi_i + 0.27C_1 + 0.10C_2 + 0.16C_3 + 0.05C_4$. 
This equation represents the buyer satisfaction value for a data buyer based on the data set's characteristics and buyer's utility.

\subsection{ Data Pricing Based on the Rubinstein Model}
After defining and discussing the concepts and roles of buyer data utility and data quality evaluation, we combine these two indicators to form a comprehensive metric for assessing buyer dataset satisfaction. This metric is designed to provide data buyers with a quantitative reference, helping them to pre-evaluate whether the dataset meets their needs and expectations during the data transaction process. Based on this, the paper proposes an improved Rubinstein pricing model under incomplete information, which incorporates the evaluation indicators of seller dataset satisfaction. This section will introduce the traditional Rubinstein model, analyze its shortcomings in the field of data pricing, and discuss the improvements made to form a multi-party data pricing model when combined with buyer dataset satisfaction.

\begin{figure}[h]
    \centering
    \includegraphics[width=0.9\linewidth]{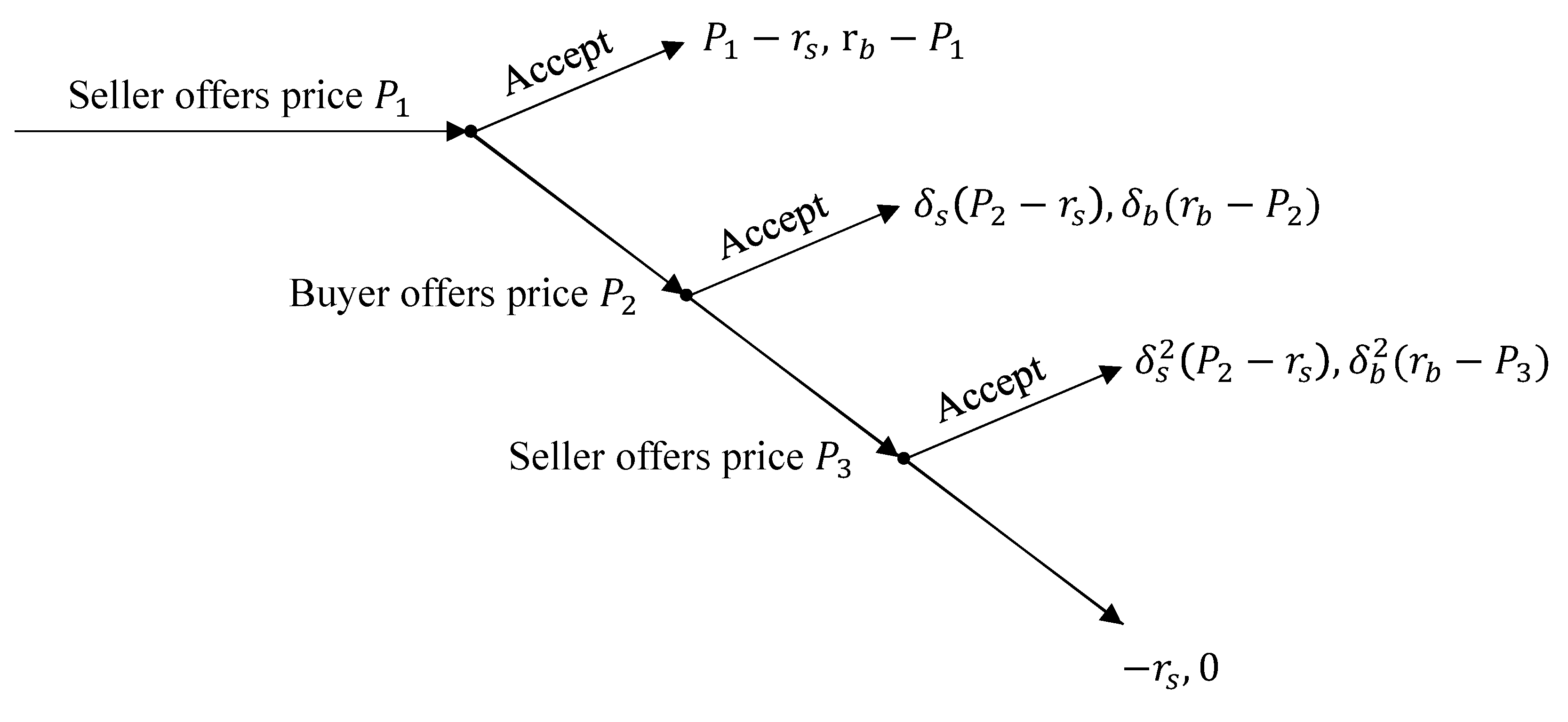}
    \caption{Three-Stage Bargaining Process Diagram.}
    \label{fig:data-bargaining}
\end{figure}

The traditional Rubinstein model, also known as the infinite game model, was proposed by Ariel Rubinstein in 1982 and is a foundational model in negotiation theory. The model aims to provide a framework for how buyers and sellers can quickly reach an agreement over time, addressing the increasing costs associated with time by using a discount factor. This paper introduces the model using the example of buyers and sellers in the traditional trading market currently being studied. Suppose that the discount factors for the buyer B and seller S during the transaction are denoted as $\delta_s$  and $\delta_b$ , respectively. The cost price of the seller $S$’s dataset is $r_s$, and the buyer’s gain from the transaction is $u$. At the beginning of the transaction, the seller offers a price $P_1$ ; if the buyer accepts, the game ends and the transaction is concluded. In this case, the seller’s payoff is $P_1-r_s$, and the buyer’s payoff is $r_b-P_1$. If the buyer rejects, the game enters the second stage. The buyer will propose a price $P_2$  that aligns with their own needs. If the seller accepts, the game ends, and the transaction is concluded. The seller’s payoff at this stage is $\delta_s (P_2-r_s)$, and the buyer’s payoff is $\delta_b (r_b-P_2)$. If the seller rejects, the game proceeds to the third stage, where the seller proposes a price $P_3$. If the buyer accepts, the game ends, and the transaction is concluded. The seller’s payoff at this stage is $\delta_s^2 (P_2-r_s)$, and the buyer’s payoff is $\delta_b^2 (r_b-P_3)$. This process continues until both parties reach a price agreement and the transaction is concluded. If no agreement is reached by the end, the seller incurs a loss of $r_s$ and time costs, while the buyer incurs time costs. Figure~\ref{fig:data-bargaining} illustrates the three-stage bargaining process of this model~\cite{ref57}.

The traditional Rubinstein model has a unique Nash equilibrium solution:
\begin{equation}
\Big( \frac{1 - \delta_b}{1 - \delta_b \delta_a}, \frac{\delta_a (1-\delta_b)}{1 - \delta_b \delta_a} \Big)
\end{equation}

The discount factor $\delta$ in the traditional model is a decreasing function of time, which implies that the bargaining process only considers the time cost of both parties. However, the traditional bargaining models assume complete information, which does not exist in real-world transaction platforms. In real-world data trading scenarios, the discount factor for both buyers and sellers is influenced not just by time costs, but also by factors such as data quality and buyer data utility.

To address this issue, the proposed model integrates the buyer's data satisfaction and data quality into the discount factor, making the discount rate a function of not only time but also the data's real utility and quality. This modification introduces a more realistic reflection of the factors affecting the bargaining process, as it considers not only the individual behavioral decisions of the buyer and seller but also incorporates the utility and quality of the data being traded.

Thus, this improvement resolves a significant limitation of traditional bargaining models—where only the personal decisions of the participants influenced the bargaining outcome—by accounting for the actual utility of the data and the quality of the data in the decision-making process. This makes the model more applicable and realistic for modern data marketplaces, where data quality, relevance, and buyer satisfaction are key factors in determining the final price and negotiation outcomes.

\subsection{Improved Rubinstein-Based Data Pricing Model}
\subsubsection{Model Assumptions}
In the data pricing and transaction model proposed in this paper, data sellers submit their dataset information and reservation price $v_0$ to the trading platform, while data buyers provide their model tasks or dataset requirements along with a budget $u$. The trading platform then forms seller alliances $S=\{S_1,S_2,...,S_n\}$ by matching sellers based on demand, and similarly, forms buyer alliances $B=\{B_1,B_2,...,B_m\}$.

The model presented in this paper is based on the following assumptions:

\begin{itemize}
\item \textit{No Collusion Among Sellers:} Data sellers are assumed to act independently without collusion.
\item \textit{Private Information:} The data buyer and seller are unaware of each other’s private information. However, within the data buyer alliance, each buyer knows their own budget $u_i$.
\item \textit{Seller’s Minimum Acceptance Price:} Based on prior research, data buyers are aware of the minimum acceptable price margin $\alpha_i$  for each seller $S_i$ and the probability $p$ that the seller will adopt a high pricing strategy.
\item \textit{Economic Maximization:} Both buyers and sellers aim to maximize their economic gains and make decisions based solely on their individual financial interests.
\item \textit{Discount Factor Based on Buyer Satisfaction:} The data buyer’s discount factor is influenced by their satisfaction with the seller’s dataset. This factor represents the degree of demand the buyer has for the data seller’s dataset. The higher the buyer’s satisfaction with the dataset, the more likely they are to engage in a transaction, as the data is expected to generate greater economic benefits. We assume that the discount factor $\delta_b$  follows a logistic decrease function with parameter $k$, as expressed by the following formula:
\begin{equation}
\delta_{b_i} = 1- \frac{1}{1+e^{-k I_{B_i}}}.
\end{equation}

\item \textit{Formation of Buyer Alliances and Joint Purchasing:} In the trading platform, data buyers form alliances and engage in joint purchasing. The needs for dataset types and features may vary among buyers. Therefore, each data buyer $B_i$ has a discount factor $\delta_{b_i}$  for the dataset provided by data seller $S_j$. The overall discount factor for the entire buyer alliance is represented as $\delta=\{ \delta_{b_1}, \delta_{b_2}, \cdots, \delta_{b_m} \}$. The collective buyer discount factor for the seller $S_j$  is weighted based on the budget share of each buyer in the alliance, as shown in the following equation:
\begin{equation}
\delta_b = \sum_{i=1}^m  \frac{u_i}{\sum_{j=1}^m u_j} \delta_{bi}.
\end{equation}

\item \textit{Effect of Platform Mediation:} The involvement of the trading platform reduces the information asymmetry between buyers and sellers, leading to an increase in the buyer’s discount factor, meaning the buyers become more patient. This adjusted discount factor is represented as $\delta_{\eta_b}= \delta_b (1+\eta)$, where $0<\eta<1$, and $\eta$ represents the level of information disclosure by the platform. The value of $\eta$ is influenced by the transparency of the trading platform, with higher transparency resulting in a higher $\eta$.

\item \textit{Impact of Higher Quality Data Needs:} Data buyers $B_i$ who require higher quality or more specifically categorized datasets must provide larger budgets, which in turn increases their influence on the overall buyer discount factor. In contrast, buyers with lower demands contribute smaller budgets, resulting in a lower impact on the collective discount factor. The total budget $r_b$  for the buyer alliance in relation to the data seller $S_j$  is determined by the weighted utility of each buyer for the dataset provided by the seller, as expressed in the following equation:
\begin{equation}
r_b = \sum_{i=1}^{m} \frac{\xi_{i,j}}{\sum_{j=1}^n \xi_{i,j}} u_i.
\end{equation}
\end{itemize}

\subsubsection{Dataset Pricing Based on Quality Evaluation}
The dataset submitted by the data seller is collected and provided to the trading platform. The seller's reservation price $r_0$  should cover both the cost of collecting, processing, and categorizing the data, denoted as $v_1$, as well as the minimum profit $v_2$  the seller requires to sustain operations. Therefore, the reservation price is expressed as:
\begin{equation}
r_0=v_1+v_2.
\end{equation}
After the data seller submits the dataset to the trading platform, the platform evaluates the dataset’s quality by using a quality assessment model. This model takes into account both the reference indicators of the data quality and the demand from data buyers on the platform. The dataset  $D$ is then assigned a quality score. The quality score $I_D$  is calculated as:
\begin{equation}
 I_D = C = \sum_{i=1}^4 W_i C_i
\end{equation}
where $W_i$ represents the weight assigned to each quality criterion $C_i$, and the summation reflects the combined impact of all quality indicators.

The quality score $I_D$  of the seller's dataset $D$ is calculated without considering the buyer's data utility indicators in this case. According to the formula from the previous section, the weight parameters $W_i$ are as follows: $W=(W_1,W_2,W_3,W_4)=(0.55, 0.13, 0.26, 0.06)$.

After the quality evaluation of the seller's dataset $D$, the dataset price $v_q$  is determined by:
\begin{equation}
r_s = r_0  I_D = r_0 (0.55C_1+0.13C_2+0.26C_3+0.06C_4 ).
\end{equation}
The price $r_s$ of the seller's dataset  $D$, after quality evaluation, becomes the seller's reservation price, which will be used in the bargaining game with the buyer.

\subsubsection{Bargaining Process}
In the trading platform, the bargaining between the buyer and seller occurs under conditions of incomplete information, where the data buyer is unaware of the seller's pricing strategy. In the infinite-period Rubinstein bargaining model defined in this paper, the trading platform assigns a price $r_s$ to the dataset after quality evaluation. Under conditions of incomplete information, the data buyer does not know the seller’s reservation price $r_s$ , but from prior research, it is known that the lowest price the seller $S_i$  will accept is $(1+\alpha_i)r_s$, and the prior probability that the buyer will face a high asking price from $S_i$ is $p_1$. In subsequent rounds of bargaining, the buyer updates the probability, adjusting it to the posterior probabilities $p=\{p_2 , p_3 , ... ,p_j\}$, where $j$ denotes the round of the negotiation. If the final transaction fails, the buyer and seller both earn zero profit. Both parties will strive to reach an agreement to maximize their respective gains. Once an agreement is reached, the trading platform will collect a commission from the final transaction amount, with the commission rate denoted as $\tau$. For simplicity, we assume $\tau=0$.

\textbf{Stage One:} The data seller proposes an initial price $P_1$. Under incomplete information, the buyer alliance is unaware of the seller's reservation price $r_s$. If the data buyer accepts the seller’s offer, the seller’s actual profit $IS_1$ is $P_1-r_s$, and the buyer alliance’s profit $IB_1$  is:
\begin{equation}
p_1 (r_b-P_1 ) + (1-p_1 )(r_b-(1+\alpha) r_s.
\end{equation}
where $1-p_1$  represents the probability that the buyer alliance guesses the seller’s asking price is low. If the buyer alliance rejects the seller’s offer, the bargaining moves to the second stage.

\textbf{Stage Two:} The buyer alliance proposes a price $P_2$. In this stage, due to the increased negotiation time cost and potential loss of benefits, both parties incur higher negotiation costs, leading to a reduction in their respective profits. If the data seller accepts the buyer alliance’s offer, the seller’s actual profit $IS_2$ is given by $\alpha_s (P_2-r_s)$. For the buyer alliance, due to the presence of the trading platform, the buyer’s discount factor increases to $\delta_{\eta b}$. The buyer alliance’s profit $IB_2$ is therefore:
\begin{equation}
\delta_{\eta b} ( p_2 (r_b-P_2 ) + (1-p_2 )(r_b-(1+\alpha) r_s ) ).
\end{equation}
If the data seller rejects the buyer alliance’s price, the bargaining moves to the third stage.

\textbf{Stage Three:} The data seller proposes a price $P_3$, and at this stage, the time cost continues to increase. If the data buyer accepts the seller’s offer, the seller’s additional profit $IS_3$  is $\alpha_s^{2} (P_3-r_0)$, and the buyer alliance’s profit $IB_3$  is:
\begin{equation}
\delta_{\eta b}^2 ( p_3 (r_b-P_3 )+(1-p_3 )(r_b-(1+\alpha) r_s ) ).
\end{equation}
If the buyer alliance rejects the seller’s price, the bargaining proceeds to the next stage.

In the case of an infinite-period bargaining process, the steps outlined above will repeat. The process will end only when one party accepts the other party’s offer.

\subsubsection{Model Solution}
When considering the infinite-period bargaining problem under incomplete information, the Harsanyi transformation provides a mechanism to transform the original problem into a bargaining game with complete but imperfect information. In this study, the third stage is chosen as the starting point for backward induction to solve the problem.

In the third stage, the data seller proposes a price $P_3$ to the buyer alliance. The seller's additional profit $IS_3$ is given by $\delta_s^2 (P_3-r_s)$, while the buyer alliance’s profit $IB_3$  is: $\delta_{\eta b}^2 (p_3 (r_b-P_3)+(1-p_3)(r_b-(1+\alpha)r_s))$. Backward induction is then applied to the second stage. The additional profit for the data seller in the second stage, $IS_2$, is given by $\alpha_s (P_2-r_s)$. If the additional profit $IS_2$  generated by the offer from the buyer alliance in the second stage is smaller than the additional profit $IS_3$ obtained in the third stage, the data seller would prefer to reject the offer from the buyer alliance in the second stage, thus increasing the time cost. To minimize the time cost and maximize economic benefit for both parties in the shortest time, the buyer alliance’s bidding strategy in the second stage should ensure that $IS_2 \geq IS_3$. Therefore, the optimal strategy for the buyer alliance is to:
\begin{equation}
IS_2=IS_3.
\end{equation}
\begin{equation}
\delta_s (P_2-r_s ) = \delta_s^2 (P_3-r_s ).
\end{equation}
\begin{equation}
\label{eq:p2-rs}
P_2=r_s+\delta_s (P_3-r_s) .
\end{equation}

That is, when the buyer alliance offers $P_2$ as given in Equation (\ref{eq:p2-rs}), the data seller will accept the offer, thus terminating the bargaining process. Under this pricing, substituting $P_2$ into the buyer alliance’s additional profit $IB_2$, we get:
\begin{equation}
IB_2=r_b+ ( \alpha(p_2-1)-1) r_s- \delta_s p_2 (P_3-r_s ).
\end{equation}

Next, we continue the backward induction process to the first stage, where the data seller makes the offer. In this stage, the data seller’s additional profit is $IS_1 = P_1-r_s$, and the buyer alliance’s profit is $IB_1=r_b-P_1$. In this stage, if the price $P_1$  offered by the data seller results in the buyer alliance’s profit $IB_1$ being lower than the profit $IB_2$  obtained in the second stage, the buyer alliance will opt to proceed to the second stage of the bargaining process, thus increasing the time cost. Therefore, for the data seller, the bidding strategy should ensure that the buyer alliance’s profit $IB_1$  is not lower than$IB_2$. The optimal strategy for the data seller should thus be:
$IB_1=IB_2$
\begin{equation}
\label{eq:ib1-ib2}
IB_1=IB_2.
\end{equation}

Substituting $P_2$ into Equation (\ref{eq:ib1-ib2}), we obtain:
\begin{equation}
\begin{aligned}
& p_1 P_1-\delta_s \delta_{\eta b} P_3 \\
& = ((p_1-\delta_s \delta_{\eta b})+ \alpha(p_1-1)+\delta_{\eta b} (1-\alpha (p_2-1))-1)r_s \\
& + (1- \delta_{\eta b})r_b.
\end{aligned}
\end{equation}

In the case of infinite bargaining rounds, considering that in both the first and third stages, the data seller is the one to make the initial price offer, followed by the negotiation phase between both parties, the price $P$ in these stages is the same. Thus, for the data seller, $P=P_1=P_3$. Therefore, the equilibrium price $P$ should be determined by the following formula:
\begin{equation}
\begin{aligned}
\label{eq:p-rs-frac}
P= r_s + \frac{(1-\delta_{\eta b} ) r_b+(\alpha(p_1-1)+\delta_{\eta b} (1-\alpha(p_2-1))-1) r_s} {p_1-\delta_s \delta_{\eta b} }
\end{aligned}
\end{equation}
When $P$ is as given in Equation (\ref{eq:p-rs-frac}), both the data seller and the buyer alliance reach an agreement, and this price is considered the equilibrium price for both parties.

\section{Experimental Evaluation}
\label{sec:experiment}
\subsection{Dataset}
The 20 Newsgroups dataset is used for experimental validation in this study. This dataset is a standard benchmark for text classification and text mining tasks. It consists of approximately 20,000 newsgroup documents, categorized into 20 distinct newsgroups, each representing a different news topic. The topics include areas such as politics, religion, sports, technology, and more. In the simulation, 10 different newsgroups are randomly selected from the dataset to form the pricing dataset.

\subsection{Experimental Setup}
The experiments in this study are conducted using Python 3.7, with the Scikit-learn machine learning framework. The experiments are run on a primary hardware setup consisting of an Intel Xeon 5218 2.30 GHz CPU and an NVIDIA V100 GPU.

\subsubsection{Trading Scenarios}
In the simulation validation of the model, the following divisions of datasets are made to verify the pricing model for sellers with exclusive datasets:

\textbf{Seller Trading Assumption 1}:
In the trading market, four data sellers $S=\{S_1, S_2, S_3, S_4\}$ and four data buyers $B=\{B_1, B_2, B_3, B_4\}$ engage in bargaining according to the market matching of supply and demand. It is assumed that data seller $S_1$ exclusively owns the data corresponding to the first label in the dataset, such as the "sci.space" category. The entire dataset is divided among the four sellers in a 3:3:2:2 ratio.

\textbf{Seller Trading Assumption 2}:
In the trading market, four data sellers $S=\{S_1, S_2, S_3, S_4\}$ and four data buyers $B=\{B_1, B_2, B_3, B_4\}$ engage in bargaining according to the market matching of supply and demand. It is assumed that data seller $S_1$ exclusively owns the data corresponding to the first label in the dataset, while the remaining datasets are distributed among three data sellers in a 4:3:3 ratio.

\textbf{Seller Trading Assumption 3}:
In the trading market, four data sellers $S=\{S_1, S_2, S_3, S_4\}$ and four data buyers $B=\{B_1, B_2, B_3, B_4\}$ engage in bargaining according to the market matching of supply and demand. In this case, it is assumed that the entire dataset is evenly distributed among all the data sellers.

In the above three trading scenarios, the four data buyers are assumed to have the following two types of demand scenarios:

\textbf{Buyer Demand Assumption 1}: 
The data purchase demand of the first buyer $B_1$ is limited to the dataset exclusively owned by data seller $S_1$.
The data purchase demand of the second buyer $B_2$ is for the entire dataset, covering all the recognized types.
The data purchase demand of the third buyer $B_3$ is for datasets excluding the one exclusively owned by data seller $S_1$.
The data purchase demand of the fourth buyer $B_4$ is for datasets including the three labels of the dataset exclusively owned by $S_1$.
The models required by buyers $B_1$ and $B_2$ are Support Vector Machines (SVM), while buyers $B_3$ and $B_4$ require Logistic Regression and K-Nearest Neighbors (KNN) models, respectively.

\textbf{Buyer Demand Assumption 2}:
The data purchase demand of all four buyers is for the entire dataset, covering all the recognized types.
The models required by buyers $B_1$ and $B_2$ are Support Vector Machines (SVM), while buyers $B_3$ and $B_4$ require Logistic Regression and K-Nearest Neighbors (KNN) models, respectively.

The purpose of these assumptions is to verify the impact of data monopoly on pricing in the current data market.
In Trading Scenario 1, where data seller $S_1$ monopolizes one category of the dataset and has the same total dataset amount as data seller $S_2$, and data sellers $S_3$ and $S_4$ have identical total dataset amounts with the same dataset categories, this scenario forms a contrast to verify the impact of the exclusive dataset on its bargaining power.
In Trading Scenario 2, data seller $S_1$ monopolizes one category of the dataset and owns only that category, verifying the effect of monopolizing a single dataset on pricing negotiations between buyers and sellers.
In Trading Scenario 3, the entire dataset is evenly divided into four parts, with each seller owning an equal amount and category of dataset. This scenario is compared with Trading Scenarios 1 and 2 to investigate the impact of data monopoly on the negotiation process.
In Buyer Demand Scenario 1, by assuming different dataset demands from buyers, the aim is to simulate real-world pricing scenarios, examining how dataset quality and the price of exclusive datasets impact the buyers' utility and satisfaction metrics.
In Buyer Demand Scenario 2, where all buyers require the full dataset, this is to contrast the effect of different buyer demands in Buyer Demand Scenario 1.

\subsubsection{Discount Factors for Buyers and Sellers}
In the pricing model proposed in this paper, we assume that the discount factor of the data seller, $\delta_s$, follows a uniform distribution between 0 and 1, i.e., $\delta_s \sim U(0,1)$. In real-world trading markets, the seller often holds an information asymmetry advantage. Therefore, in the model validation, we assume that $\delta_s > \delta_{\eta b}$.

For data buyers, the discount factor $\delta_b$ is modeled using a logistic decay function, expressed as:
\begin{equation}
\delta_{b_i} = 1 - \frac{1}{ 1+e^{-k (I_{B_i} )} }.
\end{equation}
Where $k$ is a constant parameter, and its values are selected from the set $\{7, 10, 15, 20\}$, and $I_{B_i}$ represents the satisfaction level of the $i$-th buyer with the dataset.

\begin{figure}[h]
    \centering
    \includegraphics[width=0.9\linewidth]{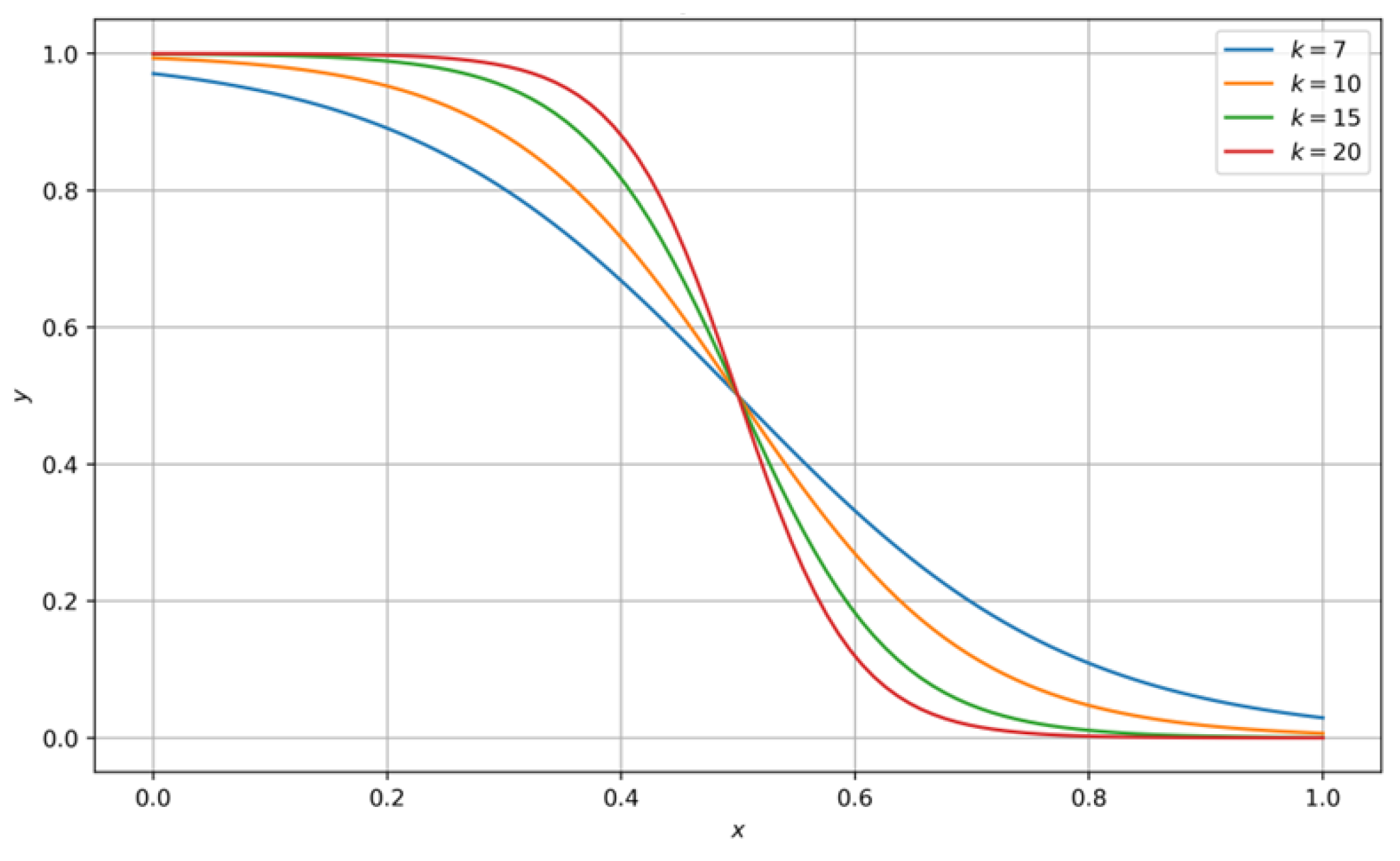}
    \caption{Logistic decay function.}
    \label{fig:Logistic}
\end{figure}

The function is plotted in Figure~\ref{fig:Logistic}, and it closely approximates the behavior of buyers in real-world markets regarding their interest in purchasing data. Specifically: 
When the buyer's dataset satisfaction $I_{B_i}$ is less than 0.5, the discount factor $\delta_b$ tends to be higher, meaning the buyer's patience is relatively high, and their interest in the dataset is low.
As the satisfaction level $I_{B_i}$ increases, the buyer's patience decreases, and the discount factor $\delta_b$ drops quickly, indicating that the buyer is more eager to obtain the dataset.
In the experiments conducted later, the value of $k$ is randomly selected from the set $\{7, 10, 15, 20\}$.

\subsubsection{Other Parameters of the Pricing Model}
In the experiments, several parameters related to quality evaluation and buyer/seller behaviors are set as follows:
\begin{itemize}
\item \textit{Quality Evaluation Parameters:} 
The value of each individual quality indicator is randomly selected from the set $\{1.2, 1, 0.6, 0.4, 0.2\}$.

\item \textit{Buyer's Price Expectation:} 
The probability that the buyer alliance guesses the data seller will demand a high price is set to $p_1 = 0.9$.
The posterior probability $p_2$ is uniformly distributed between $0$ and $p_1$, i.e., $p_2 \sim U(0, p_1)$.

\item \textit{Buyer Budget Estimation:} 
The budget estimation for each data buyer $B_i$ is uniformly distributed between 600 and 1000, i.e., $r_{B_i} \sim U(600, 1000)$.

\item \textit{Seller Reserve Price:} 
The reserve price of each data seller $S_i$ is uniformly distributed between 200 and 400, i.e., $r_{S_i} \sim U(200, 400)$.

\item \textit{Seller's Historical Minimum Profit Margin:}
The seller's historical minimum profit margin, $\alpha_{S_i}$, is uniformly distributed between 0 and 1, i.e., $\alpha_{S_i} \sim U(0, 1)$.

\item \textit{Platform Information Disclosure:}
Due to the trading platform’s role in reducing information asymmetry by providing data sellers’ information, the level of information disclosure, represented by parameter $\eta$, increases the buyer alliance's discount factor. Thus, the buyer alliance's discount factor is adjusted as $\delta_{\eta b} = (1 + \eta) \delta_b$.
\end{itemize}

These parameter settings help simulate the realistic conditions of a data marketplace where both buyers and sellers have specific expectations, budgets, and strategies shaped by their historical data and platform information.

\subsection{Experimental Results}
Tables~\ref{tab:trade1}, \ref{tab:trade2}, and \ref{tab:trade3} present the pricing outcomes under three different seller transaction assumptions. Specifically, these tables show the reserve prices generated through uniform distribution, the equilibrium prices determined after bargaining, and the additional profits for both parties, calculated by taking the difference between the pricing results and the reserve prices.

\begin{table*}[h]
  \centering
  \small
  \caption{Pricing Model Validation under Seller Transaction Assumption 1}
   \label{tab:trade1}
\begin{tabular}{ccccc}
\hline
\textbf{Buyer Demand} & \textbf{Participants} & \textbf{Reserve Price} & \textbf{Pricing Outcome} & \textbf{Additional Profit} \\ \hline
\multirow{8}{*}{Assumption 1} & $S_1$ & 365.58	& 1497.08	& 1131.50 \\
                                            & $S_2$ & 344.80	& 392.42	& 47.62 \\
                                            & $S_3$ & 243.66	& 306.79	& 63.13 \\
                                            & $S_4$ & 252.99	& 319.38	& 66.39 \\
                                            & $B_1$ & 773.09	& 743.26	& 29.82 \\
                                            & $B_2$ & 841.42	& 585.50	& 255.92 \\
                                            & $B_3$ & 967.61	& 569.83	& 397.77 \\
                                            & $B_4$ & 811.51	& 617.09	& 194.41 \\
\hline
\multirow{8}{*}{Assumption 2} & $S_1$ & 365.58	& 949.51	& 583.93 \\
                                            & $S_2$ & 344.80	& 675.69	& 330.89 \\
                                            & $S_3$ & 243.66	& 745.88	& 502.22 \\
                                            & $S_4$ & 252.99	& 638.38	& 385.39 \\
                                            & $B_1$ & 773.09	& 736.52	& 36.57 \\
                                            & $B_2$ & 841.42	& 766.90	& 74.52 \\
                                            & $B_3$ & 967.61	& 753.03	& 214.58 \\
                                            & $B_4$ & 811.51	& 752.72	& 58.79 \\
\hline                                         
\end{tabular}
\end{table*}

In Table~\ref{tab:trade1}, under Seller Transaction Assumption 1, it is assumed that data sellers $S_1$  and $S_2$  each own 30\% of the total dataset, while data sellers $S_3$  and $S_4$  each own 20\%. Regardless of whether Buyer Demand Assumption 1 or Assumption 2 is applied, $S_1$  and $S_2$  own the same amount of data, but $S_1$  generates higher additional profits due to its monopoly on a particular category of data. Since $S_3$  and $S_4$ possess the same quantity and type of data, their additional profits are quite similar. Under Buyer Demand Assumption 1, the four buyers have distinct data needs, with two buyers placing higher demand on the dataset monopolized by $S_1$ . In contrast, under Buyer Demand Assumption 2, all four buyers require datasets from all categories. There is a clear income disparity between the sellers in both assumptions. Under Buyer Demand Assumption 1, buyers $B_1$ and $B_3$  assign higher utility to the dataset monopolized by $S_1$, leading them to allocate more of their budget to $S_1$. As a result, $S_1$ earns more additional profits compared to Buyer Demand Assumption 2, which in turn reduces the profits for the other sellers. Ultimately, the pricing outcomes align with market principles.

\begin{table*}[h]
  \centering
  \small
  \caption{Pricing Model Validation under Seller Transaction Assumption 2}
   \label{tab:trade2}
\begin{tabular}{ccccc}
\hline
\textbf{Buyer Demand} & \textbf{Participants} & \textbf{Reserve Price} & \textbf{Pricing Outcome} & \textbf{Additional Profit} \\ \hline
\multirow{8}{*}{Assumption 1} & $S_1$ & 343.14	& 638.99 	& 295.86 \\
                                            & $S_2$ & 348.43	& 531.69	& 183.26 \\
                                            & $S_3$ & 288.33	& 508.52	& 220.20 \\
                                            & $S_4$ & 270.62	& 496.22	& 225.59 \\
                                            & $B_1$ & 725.78	& 469.36	& 256.42 \\
                                            & $B_2$ & 924.57	& 572.01	& 352.56 \\
                                            & $B_3$ & 627.43	& 582.39	& 45.03 \\
                                            & $B_4$ & 697.43	& 551.65	& 145.77 \\
\hline
\multirow{8}{*}{Assumption 2} & $S_1$ & 343.14	& 368.01	& 25.12  \\
                                            & $S_2$ & 348.43	& 721.00	& 372.57 \\
                                            & $S_3$ & 288.33	& 657.29	& 368.96 \\
                                            & $S_4$ & 270.62	& 637.08	& 366.46 \\
                                            & $B_1$ & 725.78	& 588.60	& 137.18 \\
                                            & $B_2$ & 924.57	& 598.60	& 325.97 \\
                                            & $B_3$ & 627.43	& 583.44	& 33.98.58 \\
                                            & $B_4$ & 697.43	& 592.74	& 114.68 \\
\hline                                         
\end{tabular}
\end{table*}

In Table~\ref{tab:trade2}, under Seller Transaction Assumption 2, it is assumed that data seller $S_1$  holds a monopoly on a single category of the dataset, while the remaining datasets are distributed as follows: data seller $S_2$  owns 40\% of the total dataset, and data sellers $S_3$  and $S_4$  each own 30\%. Under Buyer Demand Assumption 1, the four buyers each have distinct data requirements, with two buyers placing higher demand on the dataset monopolized by $S_1$ , while under Buyer Demand Assumption 2, all four buyers require datasets from all categories.

In Seller Transaction Assumption 2, there is a clear income disparity for the owner of the monopolized dataset, $S_1$, under both Buyer Demand Assumption 1 and Assumption 2. Under Buyer Demand Assumption 1, $S_1$ still generates the highest additional profits, but this is significantly reduced compared to Seller Transaction Assumption 1. In contrast, under Buyer Demand Assumption 2, $S_1$ earns the least additional profits. This is because, under Assumption 1, the utility of the monopolized dataset held by $S_1$  is higher, prompting the buyer alliance to allocate more budget to $S_1$ . However, under Assumption 2, as each buyer requires datasets from all categories, the utility for the monopolized dataset decreases. This leads to lower buyer satisfaction, resulting in reduced budgets $r_{b_{s_1} }$ and a decrease in the buyer's discount factor $\delta_b$, ultimately reducing $S_1$'s additional profits. Compared to Seller Transaction Assumption 1, in this transaction scenario, the seller holding the monopolized dataset should consider increasing its reserve price or offering additional data categories to increase its additional profits.

\begin{table*}[h]
  \centering
  \small
  \caption{Pricing Model Validation under Seller Transaction Assumption 3}
  \label{tab:trade3}
\begin{tabular}{ccccc}
\hline
\textbf{Buyer Demand} & \textbf{Participants} & \textbf{Reserve Price} & \textbf{Pricing Outcome} & \textbf{Additional Profit} \\ \hline
\multirow{8}{*}{Assumption 1} & $S_1$ & 340.02	& 558.06 	& 218.04 \\
                                            & $S_2$ & 341.83	& 572.52	& 230.70 \\
                                            & $S_3$ & 327.26	& 756.93	& 429.69 \\
                                            & $S_4$ & 258.51	& 779.86	& 521.34 \\
                                            & $B_1$ & 894.42	& 665.95	& 228.46 \\
                                            & $B_2$ & 762.25	& 667.10	& 95.15 \\
                                            & $B_3$ & 763.10	& 665.87	& 97.23 \\
                                            & $B_4$ & 900.50	& 668.45	& 232.05 \\
\hline
\multirow{8}{*}{Assumption 2} & $S_1$ & 340.02	& 673.01	& 332.99  \\
                                            & $S_2$ & 341.83	& 681.74	& 339.91 \\
                                            & $S_3$ & 327.26	& 655.88	& 328.64 \\
                                            & $S_4$ & 258.51	& 659.64	& 401.13 \\
                                            & $B_1$ & 894.42	& 667.55	& 226.87 \\
                                            & $B_2$ & 762.25	& 637.54	& 124.71 \\
                                            & $B_3$ & 763.10	& 667.71	& 95.39 \\
                                            & $B_4$ & 900.50	& 697.47	& 233.02 \\
\hline                                         
\end{tabular}
\end{table*}

In Table~\ref{tab:trade3}, under Seller Transaction Assumption 3, it is assumed that each seller owns 25\% of the total dataset, and each seller holds datasets from all categories. Under Buyer Demand Assumption 1, the four buyers each have distinct data requirements, with two buyers placing higher demand on the dataset monopolized by $S_1$, while under Buyer Demand Assumption 2, all four buyers require datasets from all categories.

The experimental results show that, under Seller Transaction Assumption 3, there is no significant difference in the additional profits of each seller. Since all sellers own the same quantity and types of datasets, the factors influencing the additional profits of seller $S_I$  in this case are primarily the buyer data utility $\xi_i$  and dataset quality $C_i$. Therefore, sellers in this scenario should focus on providing higher-quality datasets to increase their additional profits. Additionally, sellers can leverage advantages such as information asymmetry to increase their patience, which will result in an increase in the seller's discount factor $\delta_s$, significantly affecting the final price. However, the intervention of the pricing platform reduces the degree of information asymmetry, leading to a decrease in the pricing result, which reduces the seller's additional profits and increases the buyer's additional profits.

The experimental results validate the effectiveness of the multi-factor comprehensive Rubinstein pricing model. The model successfully captures the pricing advantages resulting from a seller’s monopolization of datasets, thereby increasing their additional profits. It also demonstrates the impact of different buyer models and recognition demands on the pricing results.

\begin{figure}[!t]
    \centering
    \includegraphics[width=0.8\linewidth]{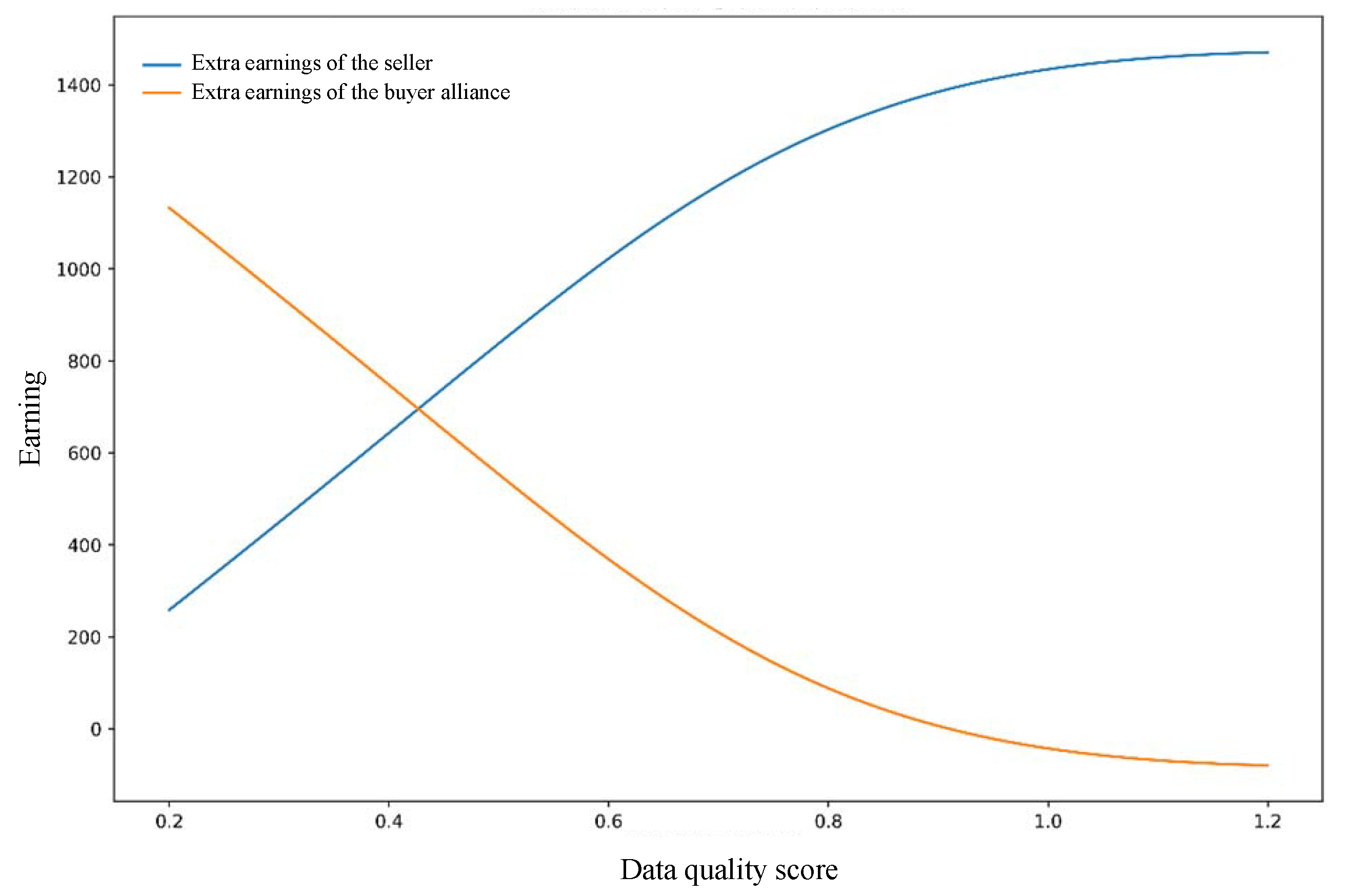}
    \caption{Data Quality Score and Multi-Party Additional Earnings Curve.}
     \label{fig:extra-earnings}
\end{figure}

Figure~\ref{fig:extra-earnings} illustrates the relationship between the additional profits of seller $S_1$  and the buyer alliance, and the changes in data quality scores under Seller Transaction Assumption 1. As shown in the figure, the seller's additional profits exhibit a positive correlation with the data quality score, while the buyer alliance's additional profits are negatively correlated with the posterior probability $p_2$. In the pricing process, higher data quality leads to higher prices for the seller's datasets $v_q$, which in turn increases the buyer alliance's dataset satisfaction. This results in a higher discount factor for the buyer alliance, and as a result, the equilibrium price increases as the quality of the dataset improves. Sellers, aiming to maximize their additional profits, are incentivized to offer higher-quality datasets, and the buyer alliance, in turn, receives more satisfactory datasets, thereby validating the model's rationality.

\begin{figure}[!t]
    \centering
    \includegraphics[width=0.8\linewidth]{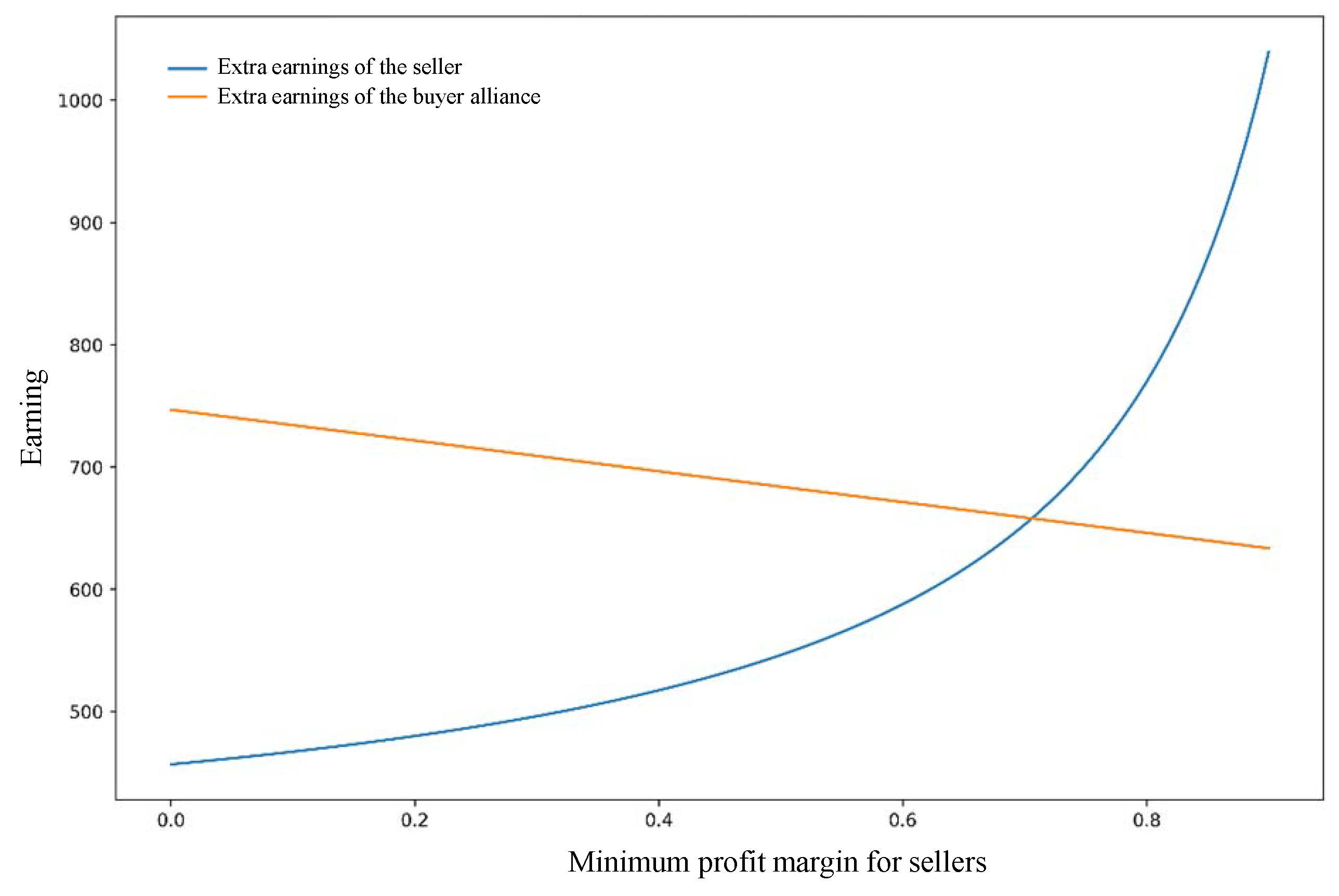}
    \caption{Seller's Minimum Profit Margin $\alpha$ and Multi-Party Additional Earnings Curve.}
    \label{fig:Minimum-profit-margin}
\end{figure}

Figure~\ref{fig:Minimum-profit-margin} depicts the relationship between the additional profits of seller $S_1$ and the buyer alliance, and the variation in seller $S_1$'s minimum profit margin $\alpha$ under Seller Transaction Assumption 1. The figure shows that the seller's additional profits are positively correlated with the minimum profit margin, while the buyer alliance's additional profits decrease as the minimum profit margin increases. This occurs because the buyer alliance can learn the minimum acceptable profit margin of the seller through prior market investigations. When the seller’s acceptable minimum profit margin is high, the buyer alliance, in an effort to quickly finalize the transaction, will raise their bid during the pricing process, thereby increasing the seller's additional profits. For the seller, to maximize additional profits, they should aim to maintain a higher profit margin in each transaction, although this may reduce the number of successful transactions, ultimately lowering the seller's total earnings.

\begin{figure}[!t]
    \centering
    \includegraphics[width=0.8\linewidth]{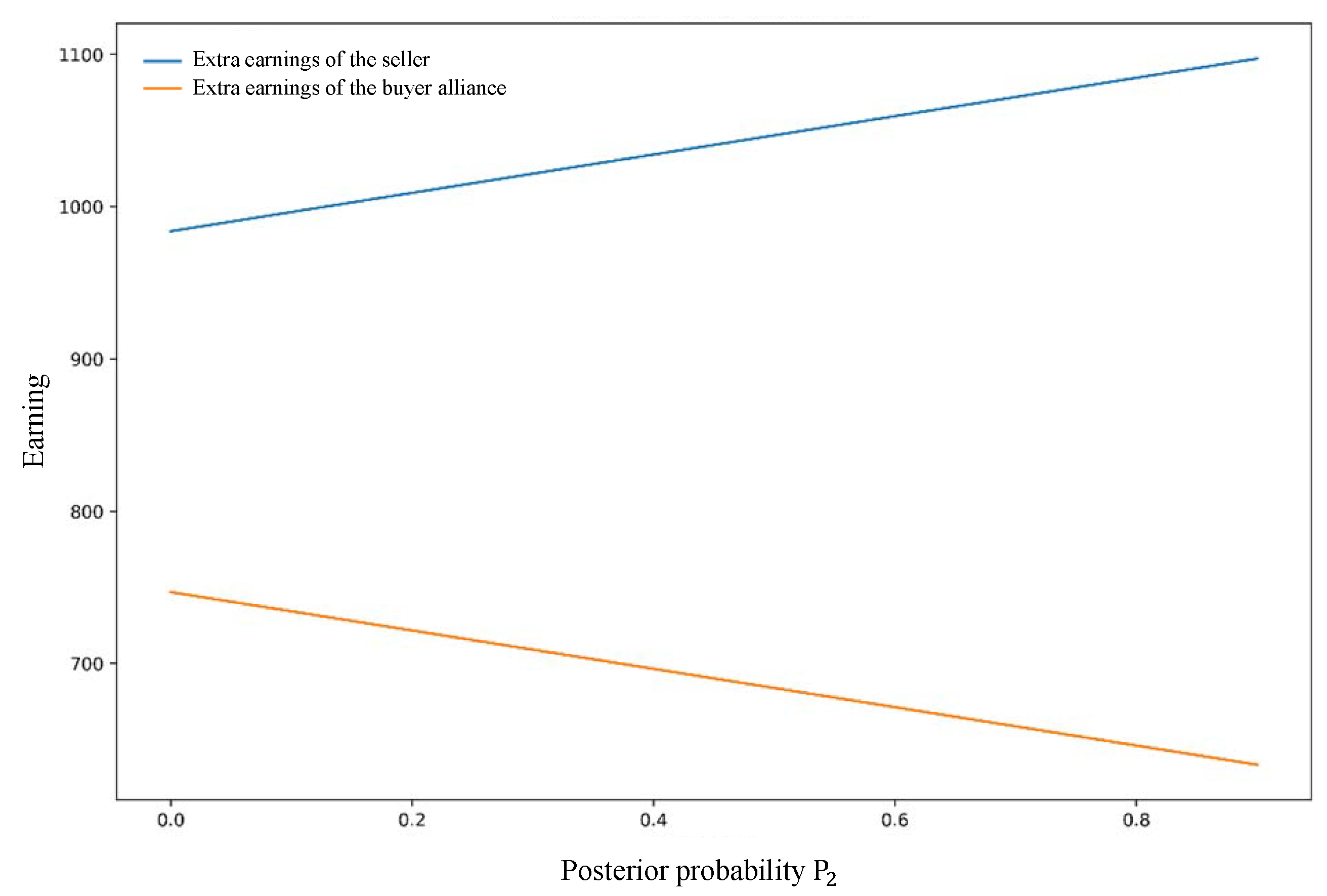}
    \caption{Posterior Probability $p_2$ and Multi-Party Additional Earnings Curve.}
    \label{fig:Posterior-probability}
\end{figure}

Figure~\ref{fig:Posterior-probability} illustrates the relationship between the additional profits of seller $S_1$ and the buyer alliance, and the variation in the posterior probability $p_2$ under Seller Transaction Assumption 1. As shown in the figure, the seller's additional profits increase as the posterior probability $p_2$ rises, while the buyer alliance's additional profits are negatively correlated with $p_2$. This occurs because when both the buyer alliance and the seller have a high willingness to engage in a transaction, considering factors like long-term cooperation and future impacts, they aim to quickly reach an agreement. In this context, the buyer alliance expects a lower posterior probability, which leads to the seller offering a lower price to the buyer alliance.

\begin{figure}[!t]
    \centering
    \includegraphics[width=0.8\linewidth]{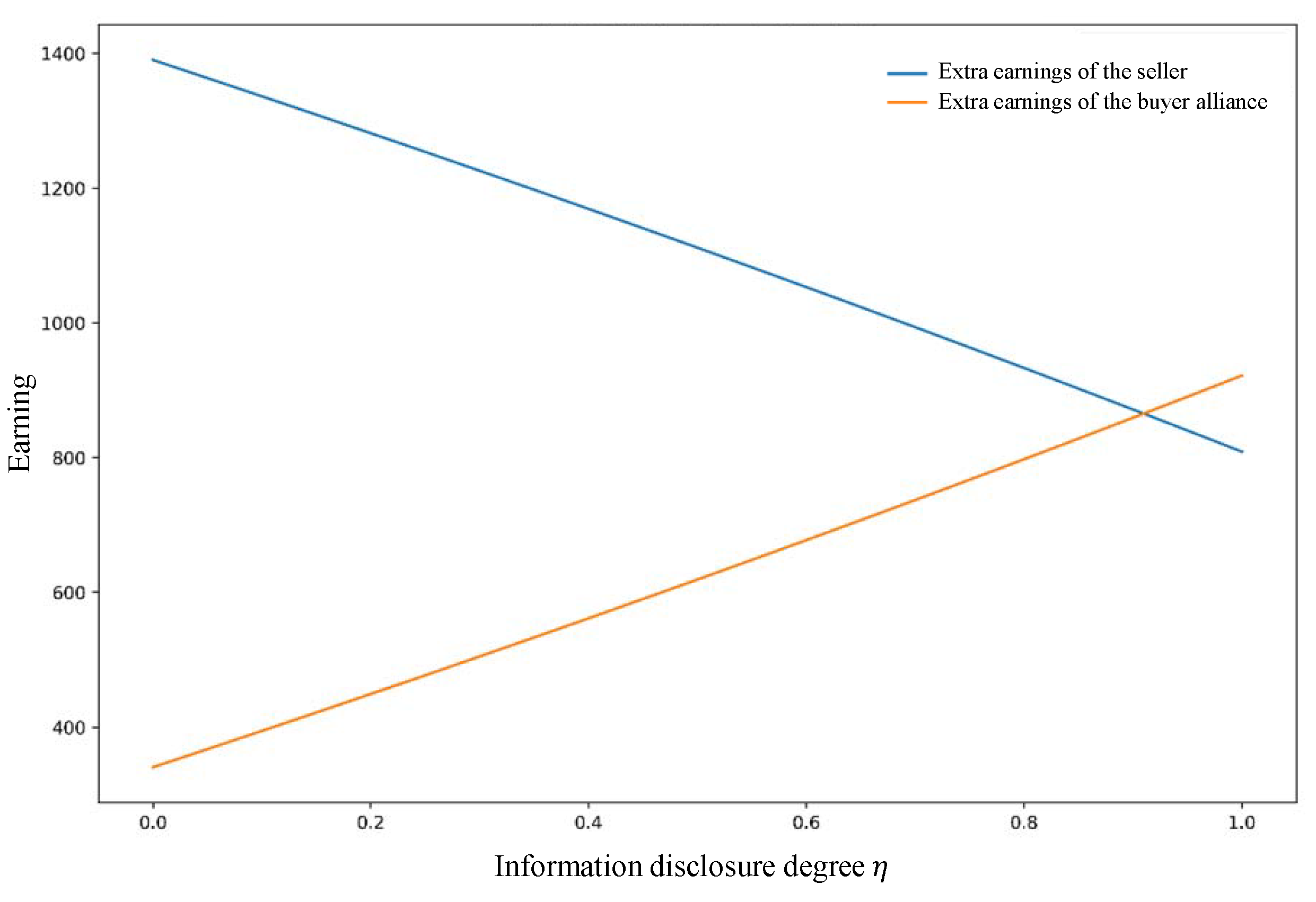}
    \caption{Information Disclosure Degree $\eta$ and Multi-Party Additional Earnings Curve.}
    \label{fig:Information-disclosure}
\end{figure}

 Figure~\ref{fig:Information-disclosure} presents the relationship between the additional profits of seller $S_1$ and the buyer alliance, and the variation in the information disclosure level $\eta$ under Seller Transaction Assumption 1. The figure demonstrates that the seller's additional profits are negatively correlated with the information disclosure level $\eta$, while the buyer alliance's additional profits are positively correlated with $\eta$. As the transparency of the trading platform increases, the buyer alliance gains more information and can better predict the strategies that the seller will adopt, thus becoming more patient with the transaction. This results in a lower equilibrium price and higher additional profits for the buyer alliance, while the seller’s additional profits decrease. In pricing, the platform should carefully consider the actual circumstances of both parties to determine the level of information transparency, ensuring a smooth transaction process for both sides.

\section{Conclusion and Future Work}
\label{sec:conclusion}
This paper primarily discusses the shortcomings present in data pricing scenarios involving multiple sellers and buyers, particularly the issues not covered by existing pricing models, such as seller-monopolized datasets, the narrow consideration of model factors, and insufficient consideration of buyer demand. To address these problems, this paper proposes a multi-factor data pricing model based on the Rubinstein model. The model introduces buyer data utility indicators, a data quality assessment model, and buyer dataset satisfaction metrics to assess the demand level of buyer $B_i$ for dataset $D_i$. By combining these evaluation methods with the Rubinstein pricing model, the model determines the price of the dataset.

In terms of experiments, three seller transaction assumptions and two buyer transaction assumptions were tested, and a thorough comparison and analysis were conducted on the impact of seller  $S_i$  monopolizing datasets and varying buyer demands on pricing. Subsequently, the effects of data quality scores, seller minimum profit margin $\alpha$, posterior probability $p_2$, and information disclosure level $\eta$ on pricing results were discussed and analyzed. The experimental results show that our proposed multi-factor comprehensive multi-party data pricing model can effectively solve the data pricing problems in the above scenarios. This research expands the application market for data pricing and provides an effective solution to the complex pricing issues in real-world trading markets.

In the proposed data pricing model, buyer data utility and data quality are combined into buyer dataset satisfaction, with the buyer alliance's discount factor determined by a logistic decrement function to reflect its influence on the bargaining process. In real-world trading, the discount factors for both buyers and sellers should be related to their personal characteristics, and these discount factors can significantly affect the final equilibrium price. Future research should further explore the rationality of parameters influencing the model's equilibrium price and the effects of changes in these parameters on the equilibrium price. The model used in this paper represents only one type of pricing mechanism, and further research is needed to explore other pricing approaches.



\end{document}